\documentclass[twocolumn,times]{aastex62}
\usepackage{natbib}
\usepackage{graphics,graphicx}

\newcommand{\kms}{km\,s$^{-1}$}
\newcommand{\cii}{[\ion{C}{2}]}

\newcommand{\lfir}{$L_{\mathrm{FIR}}$}

\newcommand{\lcii}{$L_\mathrm{[CII]}$}

\newcommand{\lsun}{$L_\sun$}
\newcommand{\msun}{$M_\sun$}

\newcommand{\msunyr}{$M_\sun$\,yr$^{-1}$}

\received{2019 February 2}
\revised{2019 March 20}
\accepted{2019 March 20}
\published{2019 April 5}

\shorttitle{400\,pc Imaging of a Quasar Host Galaxy at $z=6.6$}
\shortauthors{Venemans et al.}

\begin{document}

\title{\Large \bf 400\,pc Imaging of a Massive Quasar Host Galaxy at a Redshift of 6.6}

\correspondingauthor{Bram P.\ Venemans}
\email{venemans@mpia.de}

\author[0000-0001-9024-8322]{Bram P.\ Venemans}
\affiliation{Max-Planck Institute for Astronomy, K{\"o}nigstuhl 17, D-69117 Heidelberg, Germany}

\author[0000-0002-9838-8191]{Marcel Neeleman}
\affiliation{Max-Planck Institute for Astronomy, K{\"o}nigstuhl 17, D-69117 Heidelberg, Germany}

\author[0000-0003-4793-7880]{Fabian Walter}
\affiliation{Max-Planck Institute for Astronomy, K{\"o}nigstuhl 17, D-69117 Heidelberg, Germany}
\affiliation{Astronomy Department, California Institute of Technology, MC105-24, Pasadena, CA 91125, USA}
\affiliation{National Radio Astronomy Observatory, Pete V. Domenici Array Science Center, P.O. Box 0, Socorro, NM 87801, USA}

\author{Mladen Novak}
\affiliation{Max-Planck Institute for Astronomy, K{\"o}nigstuhl 17, D-69117 Heidelberg, Germany}

\author[0000-0002-2662-8803]{Roberto Decarli}
\affiliation{Osservatorio di Astrofisica e Scienza dello Spazio di Bologna, via Gobetti 93/3, I-40129 Bologna, Italy}

\author[0000-0002-7054-4332]{Joseph F.\ Hennawi}
\affiliation{Department of Physics, Broida Hall, University of California, Santa Barbara, CA 93106-9530, USA.}

\author[0000-0003-4996-9069]{Hans-Walter Rix}
\affiliation{Max-Planck Institute for Astronomy, K{\"o}nigstuhl 17, D-69117 Heidelberg, Germany}

\begin{abstract}

We report high spatial resolution ($\sim$0\farcs076, 410\,pc) Atacama Large Millimeter/submillimeter Array imaging of the dust
continuum and the ionized carbon line \cii\ in a luminous quasar host galaxy at $z=6.6$, 800
million years after the big bang. Based on previous studies, this galaxy hosts a
$\sim$$1\times10^9$\,\msun\ black hole and has a star-formation rate of $\sim$1500\,\msunyr.
The unprecedented high resolution of the observations reveals a complex morphology of gas
within 3\,kpc of the accreting central black hole. The gas has a high velocity dispersion
with little ordered motion along the line of sight, as would be expected from gas accretion
that has yet to settle in a disk. In addition, we find the presence of \cii\ cavities in the
gas distribution (with diameters of $\sim$0.5\,kpc), offset from the central black hole.
This unique distribution and kinematics cannot be explained by a simple model. Plausible
scenarios are that the gas is located in a truncated or warped disk, or the holes are
created by interactions with nearby galaxies or due to energy injection into the gas. In the
latter case, the energy required to form the cavities must originate from the central active
galactic nucleus, as the required energy far exceeds the energy output expected from
supernovae.  This energy input into the gas, however, does not inhibit the high rate of
star-formation. Both star-formation and black hole activity could have been triggered by
interactions with satellite galaxies; our data reveal three additional companions detected
in \cii\ emission around the quasar.

\end{abstract}

\keywords{cosmology: observations --- galaxies: high-redshift --- galaxies:
  ISM --- galaxies: star formation --- galaxies: active --- quasars: individual (VIK J030516.92--315056)}

\section{Introduction} 
\label{sec:intro}

It is expected that massive black holes in the early universe ($z\gg6$) are located in nascent massive galaxies \citep[e.g.,][]{val14,dim17}. At these high redshifts, supermassive black holes are not detected directly, but their presence is inferred as they power luminous quasars through accretion. Such luminous quasars have been found up to $z=7.5$ \citep{ban18a}, indicating that supermassive black holes were already formed at $z>7$ \citep[e.g.,][]{mor11,der14,maz17b,ban18a}. 

Using (sub-)millimeter facilities, such as the Atacama Large Millimeter/submillimeter Array (ALMA) and the IRAM NOrthern Extended Millimeter Array, the interstellar medium (ISM) of $z>6$ quasar host galaxies have now been routinely detected \citep[e.g.,][]{ber03b,wal03,wal04,mai05,wan10,wan13,ven12,ven16,ven18,wil13,wil15,dec18}. While  the global properties of the quasar hosts are now reasonably well determined \citep[e.g.,][]{dec18,ven18}, most observations barely resolve the galaxies (i.e., apparent galaxy size $\lesssim2$ times the size of the resolution of the observations). The outstanding imaging capabilities of ALMA now allow us to investigate the ISM in quasar host galaxies well below kiloparsec scales, and study the distribution and kinematics of gas and dust in extraordinary detail. 

The quasar J0305--3150 at $z=6.6$ was
discovered in 2013 \citep{ven13} and has a black hole mass of
$\sim1\times10^9$\,\msun\ \citep{ven13,der14,maz17b}. Previous ALMA observations at 0\farcs62 (3.4\,kpc at $z=6.6$) resolution of the \cii\ 158$\mu$m
emission line and the underlying dust continuum \citep{ven16} revealed
that the black hole is hosted by an ultra-luminous infrared host galaxy
(far-infrared (FIR) luminosity $L_\mathrm{FIR}$ exceeding $10^{12}$\,\lsun). Based on the dust continuum and the detection 
of CO(6--5) and CO(7--6) \citep{ven17b}, the estimated molecular gas mass in J0305--3150 is
$(2.4-18)\times10^{10}$\,\msun. This gas mass is $>$60\% of the
dynamical mass derived from the extent and width of the
\cii\ line \citep{ven16}, which has an 
FWHM of 255\,km\,s$^{-1}$. These earlier observations indicated that the \cii\ emission line spectrum shows
nonvirial motion, consistent with the presence of an outflow or companion
galaxy.

Here we present high spatial resolution (0\farcs076) ALMA observations of the host galaxy of quasar J0305--3150. 
The Letter is organized as follows. In Section~\ref{sec:obs} we provide details of the new observations. The results are discussed in Section~\ref{sec:results}. We present the spatial distribution of the gas and dust in the quasar host in Section~\ref{sec:moments} and we investigate the gas kinematics in Section~\ref{sec:kinematics}. In Section~\ref{sec:models} we introduce a model to understand the observed gas kinematics. The constraints from the model are presented in Section~\ref{sec:modelresults}. In Section~\ref{sec:cavities} we explore alternative scenarios to explain the cavities seen in the gas distribution and introduce the presence of companion \cii\ emitters in the field in Section~\ref{sec:companions}. We conclude with a summary in Section~\ref{sec:summary}. 

Throughout this Letter we adopt a Lambda cold dark-matter cosmology with a Hubble constant of $H_0 = 70$\,\kms\,Mpc$^{-1}$, a mass density of $\Omega_M = 0.3$ and a vacuum density of $\Omega_\Lambda = 0.7$, which is consistent with the latest {\it Planck} measurements \citep{pla16}. With these cosmological parameters, the age of the universe at $z=6.6$ is 810\,Myr. Far-infrared luminosities, \lfir, are computed by integrating the dust spectral energy distribution (SED) between the rest-frame wavelengths 42.5 and 122.5\,$\mu$m \citep[e.g.,][]{hel88}. For the shape of the dust SED we follow the literature and assume a dust temperature of 47\,K and a dust emissivity index of $\beta=1.6$ \citep[e.g.,][]{bee06,wan08b,wan13,wil13,ven16}. Uncertainties in the FIR luminosity quoted in this Letter represent measurement uncertainties only. The uncertainties introduced in \lfir\ due to the unknown shape of the dust SED are a factor of 2--3 \citep[see the extensive discussion in][]{ven18}.

\section{ALMA Observations}
\label{sec:obs}

To resolve the structure of the host galaxy and explore its detailed kinematics, we observed J0305--3150 between 2017 November 12 and 18 with ALMA in configuration C43-8. The number of antennas was 43 with baselines between 92 and 13,894\,m. The ALMA observations covered the redshifted \cii\ line observed at 249.6\,GHz with a single bandpass with a frequency width of 1.875\,GHz ($\sim$2250\,\kms). Three other bandpasses of 1.875\,GHz each were placed to measure the FIR continuum at observed frequencies of 252.1, 264.6, and 266.6\,GHz. Bandpass and flux calibration was performed through observations of J0522--3627. For the phase calibration, the source J0326--3243 was observed. The total observing time was 7.6\,hr, of which 3.5\,hr were on-source. 

The data were reduced using the Common Astronomy Software Application package \citep[][]{mul07}, following the standard reduction steps. The rms noise around the redshifted \cii\ line is 87\,$\mu$Jy\,beam$^{-1}$ per 30\,MHz bin (36\,\kms).
The final beam using natural weighting has a size of 0\farcs076\,$\times$\,0\farcs071, which corresponds to 410\,pc\,$\times$\,385\,pc at $z=6.6$. With this resolution, we can resolve structures on scales of $\sim$400\,pc in the quasar host galaxy and look for possible interplay between the accreting black hole and its host. For a black hole mass of $\sim$$10^9$\,\msun\ and with the gas having a velocity dispersion of $\sim$110\,km\,s$^{-1}$ (based on the FWHM of 255\,km\,s$^{-1}$), the region where the black hole dominates the gravitational potential, the so-called black hole sphere of influence, has a radius of $\sim$355\,pc. Thus the achieved spatial resolution of the observations in principle allow us to probe scales similar to the black hole sphere of influence in order to search for kinematic signatures of the central black hole.

\section{Results}
\label{sec:results}

\begin{figure*}
\begin{center}
\includegraphics[width=\textwidth]{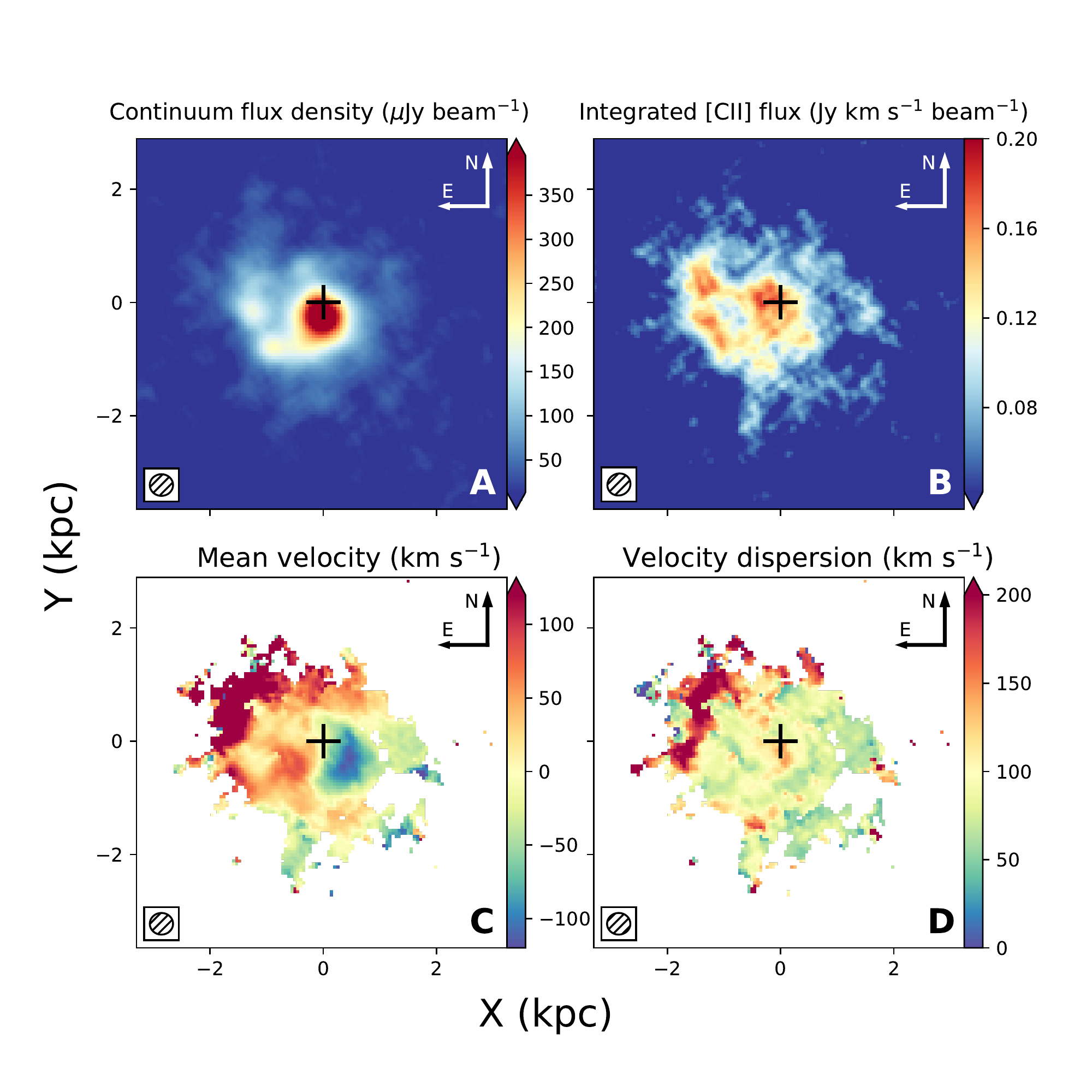}
\caption{Dust continuum and \cii\ intensity
maps (top row) and \cii\ velocity (moment 1) and dispersion
(moment 2) maps (bottom row). In each panel, the size of the beam (0\farcs076$\times$0\farcs071 in all four maps) is plotted in the bottom left corner. The cross indicates the observed NIR location of the
quasar and the size of the cross represents the astrometric uncertainty of the quasar position. {\bf (A)} Image of the dust continuum at an observed frequency of 259.3\,GHz (1974\,GHz in the rest frame). 
 {\bf (B)} Map of the integrated \cii\ flux, showing the emission from
--125\,\kms\ to +150\,\kms. {\bf (C)} The kinematics of the \cii\
line (moment 1). {\bf (D)} Map of the velocity dispersion of the \cii\
emission (moment 2).}
\label{fig:moments}
\end{center}
\end{figure*}

\subsection{Dust and \cii\ emission}
\label{sec:moments}

\begin{figure}
\begin{center}
\includegraphics[width=\columnwidth]{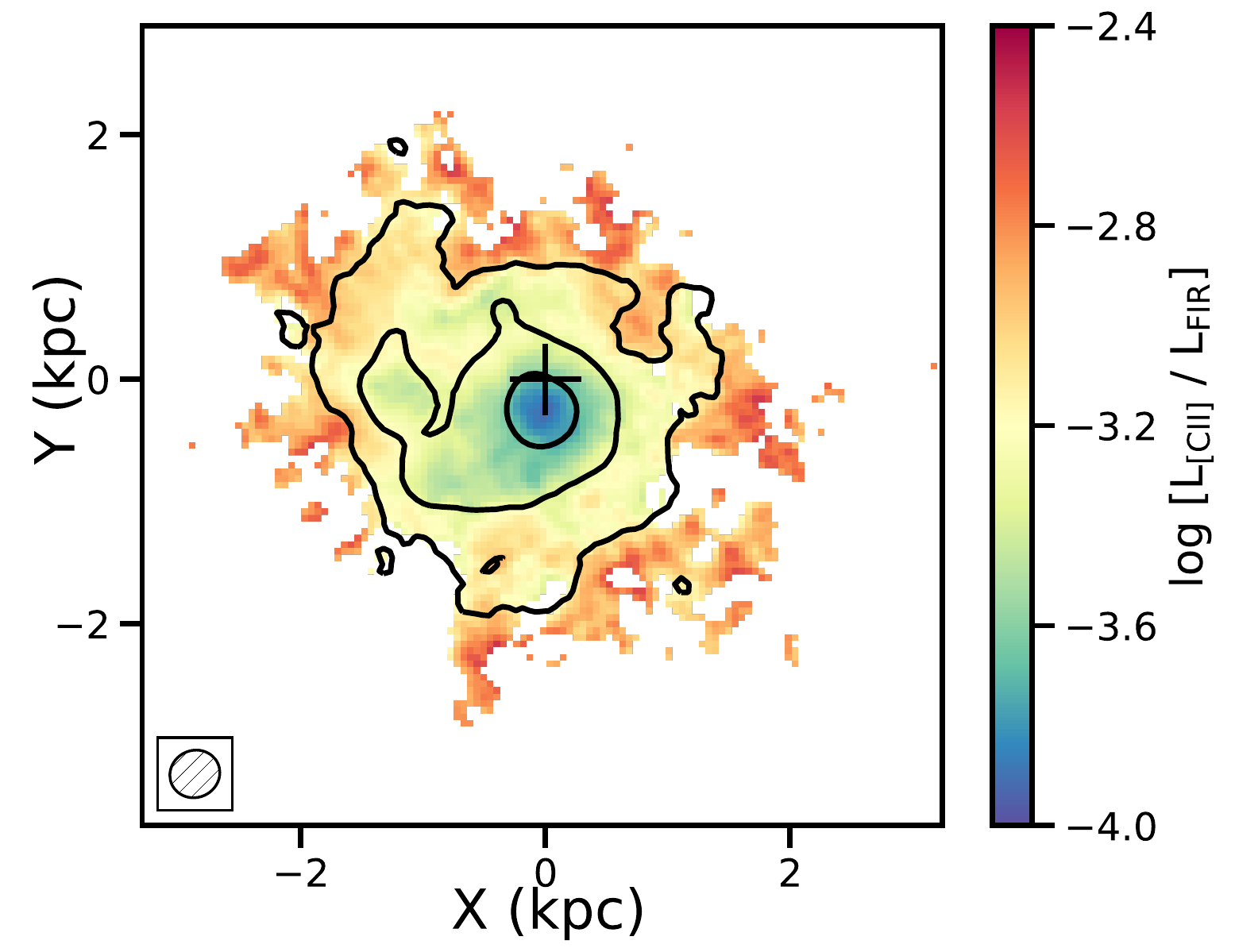}
\caption{Map of the \cii-to-FIR luminosity ratio. The \lcii/\lfir\ ratio is plotted in color, and the contours represent the dust emission at 6\,$\sigma$, 18\,$\sigma$, and 54\,$\sigma$ levels (see also Figure~\ref{fig:moments}). The lowest values in \lcii/\lfir\ coincide with the peak of the dust continuum emission and the position of the quasar. }
\label{fig:lumrat}
\end{center}
\end{figure}

In Figure~\ref{fig:moments} we show the results of the observations. The first two panels in the figure show the \cii\ and continuum maps. The bottom two panels in Figure~\ref{fig:moments} show the kinematics of the \cii\ emission. 
To match the position of the \cii\ and FIR dust emission with that of the accreting black hole, we recomputed the near-infrared (NIR) location of the quasar by correcting the positions of nearby stars with a match in the GAIA DR2 catalog \citep{gaia18}. The resulting NIR location of the quasar is 03$^\mathrm{h}$05$^\mathrm{m}$16$^\mathrm{s}\!$.919, --31$^\circ$50$^\prime$55\farcs86 with an uncertainty of 0\farcs02, which is consistent with the previously published coordinates of the quasar \citep{ven13,ven16}. The distribution of the FIR continuum
emission has a pronounced peak that is coincident with the
NIR location of the quasar. 

We measured the total flux of the quasar host galaxy in an aperture with a radius of 0\farcs75. The total continuum flux density is 5.34$\pm$0.19\,mJy around the \cii\ line, making this quasar host one of the most luminous, unlensed sources
known at high redshift \citep[e.g.,][]{ven18}. The total brightness of the \cii\ line is 5.43$\pm$0.33\,Jy\,\kms, which corresponds to a luminosity of \lcii\,=\,$(5.90\pm0.36)\times10^9$\,\lsun. We derive a systemic redshift of the quasar host by fitting a Gaussian to the \cii\ emission line, resulting in $z_\mathrm{[CII]}=6.61391\pm0.00015$, consistent with earlier measurements \citep{ven16}.

Both the continuum and the \cii\ emission are spatially resolved, and
the emission is extended over $\sim$5\,kpc. The gas distribution and
kinematics, as traced by the \cii\ emission, are highly
complex. There is a pronounced lack of \cii\ emission toward the east of the quasar. This cavity in the \cii\ emission is also seen in the dust continuum observations. In general, the continuum and \cii\ emission trace similar structures, the main difference being the bright peak in the continuum.

In Figure~\ref{fig:lumrat} we compare the \cii\ emission to that of the dust. The \lcii/\lfir\ ratio is lowest at the location of the quasar where the dust continuum peaks (\lcii/\lfir\,$\approx1.5\times10^{-4}$). This is reminiscent of the centers of local star bursts and ultraluminous infrared galaxies (ULIRGs), where low \cii-to-FIR luminosity ratios are observed in regions with high FIR surface brightness \citep[e.g.,][]{smi17,her18}. Away from the central regions, the \cii-to-FIR luminosity ratio is in the range of \lcii/\lfir\,$=(0.5-1)\times10^{-3}$. More centrally concentrated continuum emission has been observed in some other high-redshift galaxies and quasar hosts, albeit at lower spatial resolution \citep[e.g.,][]{wan13,cap15,ven16,gul18}.

\subsection{Gas Dynamics}
\label{sec:kinematics}

\begin{figure*}
\begin{center}
\includegraphics[width=\textwidth]{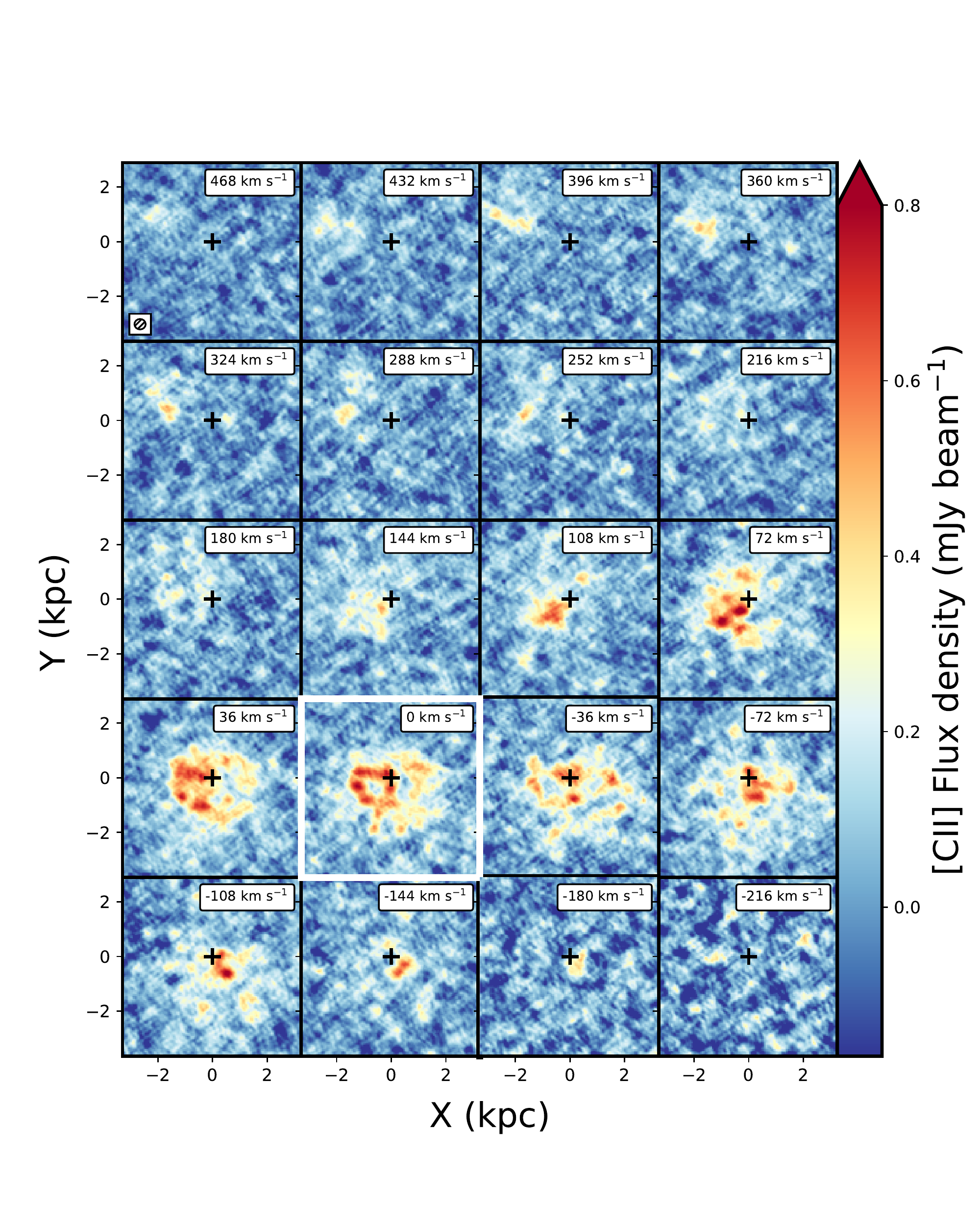}
\caption{Channel maps of the \cii\ emission. For each image the emission was averaged over 30\,MHz (36\,\kms). The cross indicates the location of the quasar (here defined as the position of the peak of the FIR continuum; see Section~\ref{sec:moments}). At the top of each image the average velocity (compared to the systemic redshift of $z=6.61391$) is indicated.}
\label{fig:channel}
\end{center}
\end{figure*}

From the mean velocity map of the \cii\ emission (third panel in Figure~\ref{fig:moments}) it is evident that the position of the accreting black hole
coincides with the kinematic center of the \cii\ emission. It
is also located at the center of the global \cii\
emission. Generally, the gas exhibits some ordered motion along the line of sight, with the gas
having positive line-of-sight velocities toward the east and negative velocities
toward the west. The highest velocity component, which was already discovered
in low-resolution data \citep{ven16}, is now clearly separated, both spatially and in frequency. It is
therefore likely that this is a companion galaxy close to the quasar
host (see also Figures~\ref{fig:channel} and \ref{fig:companions}). This will be further discussed in Section~\ref{sec:companions}. 

The velocity dispersion in the quasar host is roughly uniform
(in the range 50--100\,\kms) throughout the galaxy. The dispersion is almost equal to the projected line-of-sight velocity. At the position of the quasar, there is no sharp
increase in the velocity dispersion; the dispersion of 110\,\kms\ at that location is
not found to be higher compared to the remainder of the host galaxy.
This indicates that, with the current resolution, the
mass budget within the central $\sim$400\,pc is not dominated by
the black hole. Indeed, assuming typical dust properties \citep[dust
temperature $T_d=47$\,K and an emissivity index
$\beta=1.6$; e.g.,][]{bee06} and a gas-to-dust ratio of 70 \citep[e.g.,][]{san13},
the inferred gas mass within our central resolution element
is $6\times10^9$\,\msun\ \citep[with significant uncertainties; see, e.g., the discussion in][]{ven18}, higher than the $10^9$\,\msun\ of the black hole \citep{der14}.

To explore how the distribution of \cii\ emission changes with
line-of-sight velocity, we have averaged the emission into channels with a width of
30\,MHz (36\,\kms; see Figure~\ref{fig:channel}). There are two striking features in the
channels centered around 0\,\kms: (i) the \cii\
emission covers the whole spatial extent seen in the integrated
emission map (Figure~\ref{fig:moments}), and (ii) there are two depressions/cavities in
the \cii\ emission with diameters of $\sim$\,0.5\,kpc on either
side of the black hole. The gas with the highest velocities --- that was
already seen at positive velocities toward the northeast in the low-resolution data \citep{ven16} --- is clearly offset from the quasar host.

\begin{figure}
\begin{center}
\includegraphics[width=\columnwidth]{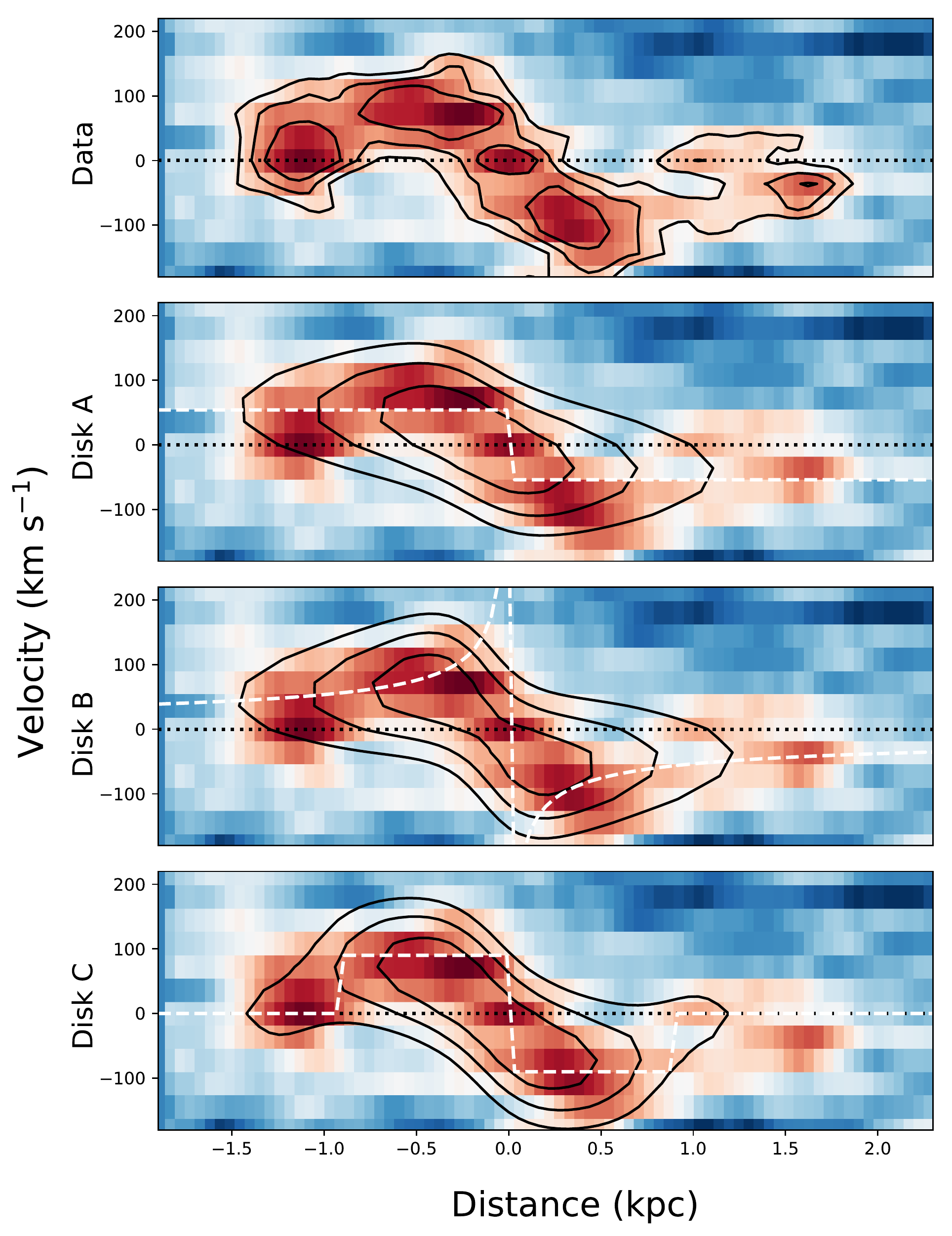}
\caption{Position--velocity diagram along the major axis, illustrating how the velocity of the gas changes as a function of distance from the black hole. The major axis is aligned approximately east to west. In each panel, the color image represents the data. In the bottom three panels, the black contours show the fit to the data of three different models. These models are described in detail in Section~\ref{sec:modelresults}. 
}
\label{fig:pv}
\end{center}
\end{figure}

We can capture some of the complexity of the kinematics by generating
a position-velocity diagram of the \cii\ emission (Figure~\ref{fig:pv}). A position-velocity diagram is a planar slice through the data cube and a useful way to visualize the complex kinematics as it shows the line-of-sight velocities of the gas as a function of distance from the center. This pseudo-longslit spectrum, oriented along the major kinematical axis through the position of
the quasar and the \cii\ cavity to the east, shows an S-like shape in the velocity curve. This implies that, in
addition to the cavity that is clearly seen toward the east, there is
a second cavity to the west, which is also seen in the channel maps at
zero velocity (Figure~\ref{fig:channel}). Interestingly, the line-of-sight velocities approach 
zero at the largest radii, consistent with the finding that the channel map
at zero velocity shows the full extent of the integrated line
emission. 

In the next Section, we will fit several models to explain the distribution and velocity of the gas.

\section{Kinematic Modeling of the \cii\ emission}
\label{sec:models}

\begin{deluxetable*}{lcccc}
\tablecaption{Model Parameters. \label{tab:model}}
\tablewidth{0pt}
\tablehead{\colhead{} & \colhead{Disk A} & \colhead{Disk B} & \colhead{Disk C} & \colhead{Rot + Jet D}}
\startdata
Velocity profile & Constant & Keplerian & Truncated & Constant \\
R.A.$^{\rm a, b}$ & 
03$^\mathrm{h}$05$^\mathrm{m}$16$^\mathrm{s}\!$.9225(1) & 
03$^\mathrm{h}$05$^\mathrm{m}$16$^\mathrm{s}\!$.9219(1) &
03$^\mathrm{h}$05$^\mathrm{m}$16$^\mathrm{s}\!$.9219(1) &
03$^\mathrm{h}$05$^\mathrm{m}$16$^\mathrm{s}\!$.9227(2) \\
decl.$^{\rm a, b}$ & 
--31$^\circ$50$^\prime$55\farcs938(2) &
--31$^\circ$50$^\prime$55\farcs939(2) &
--31$^\circ$50$^\prime$55\farcs940(1) & 
--31$^\circ$50$^\prime$55\farcs939(3) \\
Redshift$^{\rm a, b}$ & $6.61405(3)$ & $6.61396(3)$ & $6.61388(3)$ & $6.61404(3)$ \\
Inclination (deg.) & $38.0^{+2.0}_{-2.8}$ & $38.5^{+1.6}_{-2.0}$ & $34.7^{+1.8}_{-1.7}$ & --- \\ 
Position angle (deg.) & $257^{+1.4}_{-1.4}$ & $262^{+1.2}_{-1.3}$ & $271^{+1.2}_{-1.1}$ & $249^{+1.9}_{-2.0}$ \\
Central intensity (mJy\,beam$^{-1}$) & $0.78^{+0.11}_{-0.12}$ & $0.85^{+0.11}_{-0.11}$ & $0.86^{+0.13}_{-0.12}$ & $0.81^{+0.33}_{-0.31}$\\
Intensity scale radius (kpc) & $1.41^{+0.03}_{-0.03}$ & $1.34^{+0.03}_{-0.03}$ & $1.29^{+0.03}_{-0.02}$ & $0.91^{+0.02}_{-0.02}$\\
Circular velocity (\kms) & $87^{+6}_{-4}$ & $96^{+23}_{-18}$ & $157^{+8}_{-9}$ & $103^{+9}_{-5}$ \\ 
Velocity dispersion (\kms) & $99.3^{+0.9}_{-0.9}$ & $95.6^{+1.0}_{-0.98}$ & $94.6^{+0.98}_{-1.0}$ & $99.7^{+1.1}_{-1.2}$ \\ 
Velocity scale radius (kpc) & --- & $0.81^{+0.39}_{-0.28}$ & $0.98^{+0.04}_{-0.01}$ & --- \\
Jet opening angle (deg.) & --- & --- & --- & $28.6^{+2.7}_{-2.9}$ \\
Jet $z$-axis angle (deg.)$^{\rm c}$ & --- & --- & --- & $37.8^{+2.7}_{-2.9}$ \\
Jet $x$-axis angle (deg.)$^{\rm d}$ & --- & --- & --- & $-135^{+7}_{-7}$ \\
\enddata
\tablenotetext{a}{Position of the kinematic center of the model.}
\tablenotetext{b}{Number in parentheses is the uncertainty in the last number.}
\tablenotetext{c}{Angle of the jet compared to the axis of rotation.}
\tablenotetext{d}{Angle of the jet compared to the position angle of the rotation.}
\vspace{-0.5cm}
\end{deluxetable*}

\begin{figure}
\begin{center}
\includegraphics[width=\columnwidth]{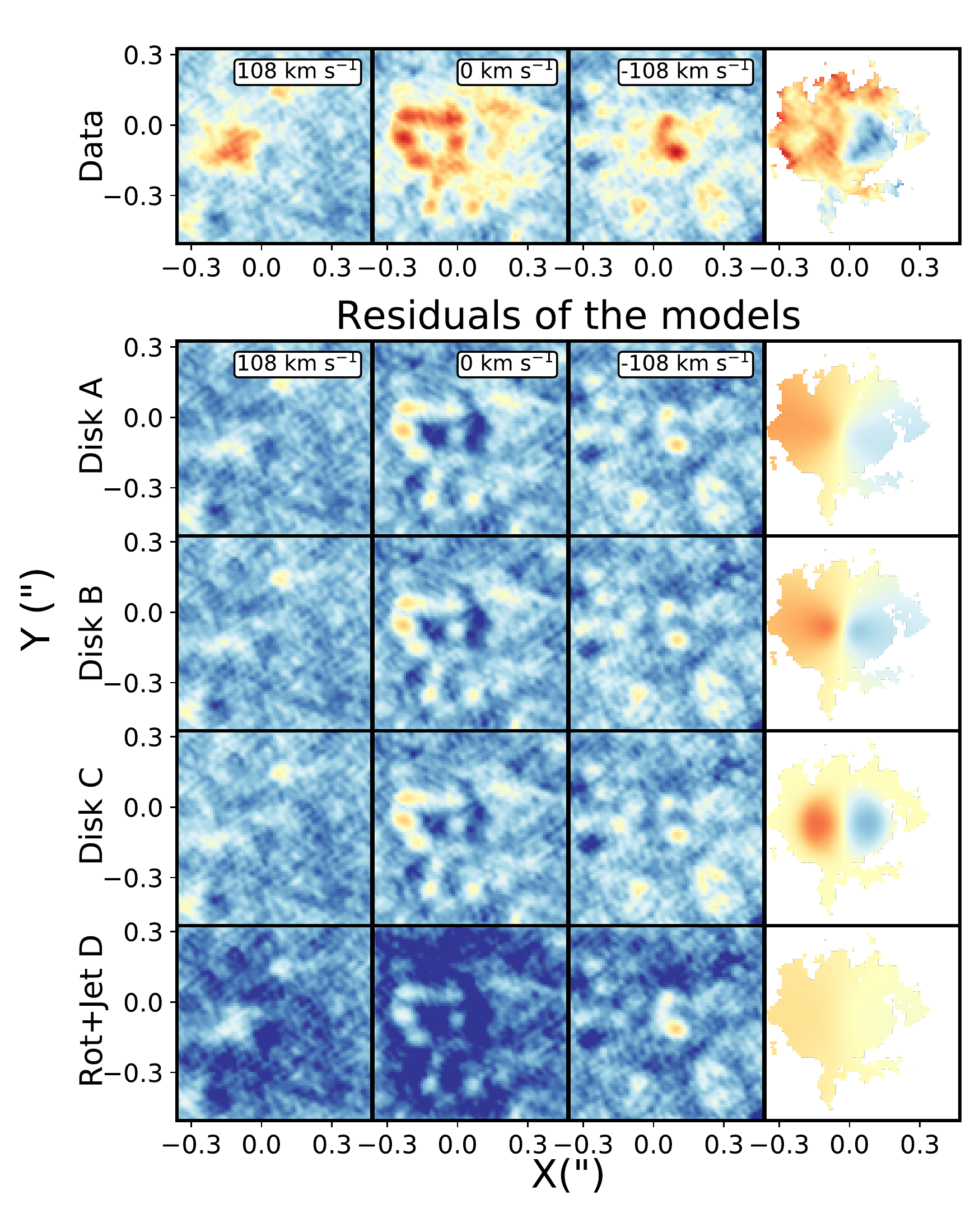}
\caption{Figure of the different models. The three
leftmost columns show three representative velocity channels of the full cube centered on the \cii\ emission. The rightmost column is the intensity-weighted velocity map. The top row is the data, whereas the remaining four rows shows the four different models described in Section~\ref{sec:modelresults}. The velocity map for the four models is only shown in those pixels that have a $3\sigma$ integrated flux density in the data. }
\label{fig:models}
\end{center}
\end{figure}

\begin{figure*}
\begin{center}
\includegraphics[width=\textwidth]{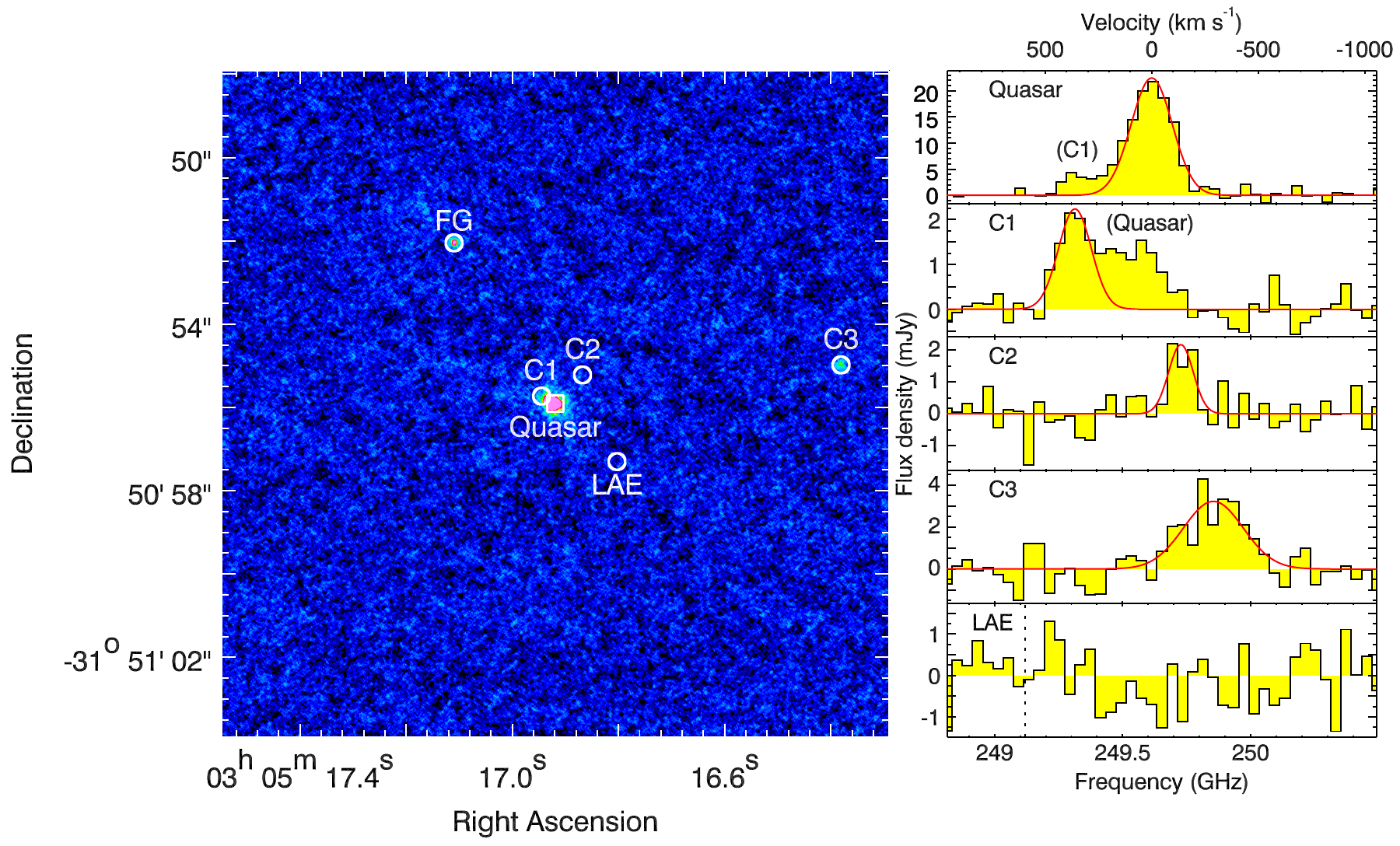}
\vspace{-0.5cm}
\caption{{\it Left:} Continuum image of the field surrounding the quasar. This image size is 16$^{\prime\prime}$\,$\times$\,16$^{\prime\prime}$ (86.5$\times$86.5\,kpc$^2$). Several objects are detected near the quasar and are marked with a circle. Three \cii\ emitting companions and a Lyman-alpha emitting galaxy \citep{far17} are located within 40\,kpc and $\sim$1000\,\kms. The ALMA spectrum of the continuum source labeled ``FG" did not reveal any emission lines. This source is likely not located near the quasar, but in the foreground. {\it Right:} Spectra of the \cii\ emitting companions and the quasar host galaxy. The characteristics of the companion galaxies are listed in Table~\ref{tab:companions}. Due to the close proximity of C1 to the quasar host, the spectra of both C1 and the quasar host are overlapping. This is indicated in the spectra.}
\label{fig:companions}
\end{center}
\end{figure*}

To better understand the spatial extent and kinematic signature of the
\cii\ emission, we have fitted a range of simple models to the
data. These models are discussed below and are summarized in Table~\ref{tab:model}. 
To fit the models, we generate a cube from the given model
parameters, which we will refer to as the model cube. This model cube
is then convolved with the ALMA primary beam to account for potential
beam smearing effects. We compare this convolved model data cube,
$M$, with the data cube, $D$, using a standard chi-squared algorithm,

\begin{equation}
{\chi}^2 = \sum_{i} (M_i - D_i)^2 / U_i^2
\end{equation}

\noindent
where $U_i$ is the uncertainty (i.e., rms
uncertainty of the pixels showing no emission in the same channel as
pixel $i$). The model cube has velocities that range from --216 to +216\,\kms. This range covers the velocities where the data show \cii\ emission but excludes the channels affected by the close companion (see Figures~\ref{fig:channel} and \ref{fig:companions}).
Since the adjacent pixels are highly correlated in
interferometric data, we do not sum up all of the pixels in our data, but
instead use a bootstrap method, whereby we randomly select $N$ pixels
from the data cube and calculate the $\chi^2$ value of these
pixels. We repeat this process until the median $\chi^2$ value remains
unchanged. Here the number of pixels, $N$, is chosen so that on
average each beam only contains a single pixel per single $\chi^2$
calculation.

To fully sample the parameter space of the model and provide more
realistic constraints on the possible values of each parameter, we
have used a Markov chain Monte Carlo (MCMC) method to sample the
parameter space. In particular, we have used the affine invariant MCMC
ensemble sampler code, \texttt{emcee} \citep{for13}. For
all parameters, we assume flat priors, and we initiate the parameters
with rough estimates from visual inspection of the data. We have
verified that these initial guesses do not affect the results of the
MCMC analysis. Results of the different models are shown in Figure~\ref{fig:models} and Table~\ref{tab:model}.

\subsection{Modeling Results}
\label{sec:modelresults}

We have modeled the \cii\ kinematics in 3D using several
simple models. The following models were considered in this analysis: (a) a thin
rotating disk with a constant velocity profile (Disk A). This is the fiducial
profile of a dark-matter-dominated disk galaxy, which normally is the model employed to explain the gas kinematics in marginally resolved $z\sim6$ quasar host galaxies \citep[e.g.,][]{wan13,ven16,sha17}. Model (b) is a thin rotating
disk with a decreasing velocity profile (Disk B). This profile would arise if
the mass of the galaxy is centrally located. Model (c) is a thin rotating disk
with a constant velocity profile up to a certain radius, after which
it decreases to zero (Disk C). This is the velocity profile of a disk that is
truncated at a certain radius, or the velocity profile of a disk that
is warped into the plane of the sky at this radius. Model (d) is an inclined
biconical jet embedded in a uniform rotating spherical gas (Rot+Jet D). This is a
simple model of an active galactic nucleus (AGN) jet accelerating and/or ionizing the
\cii\ emitting ISM. This possibility will be further discussed in Section~\ref{sec:cavities}. All of the models have a decreasing
intensity profile dependent only on the distance from the center. This
clearly is an oversimplification, as the moment 0 map shows obvious
evidence of nonexponential emission (see Figure~\ref{fig:moments}).

The number of parameters in each of the models varies from 9 to
12 (Table~\ref{tab:model}). Each model contains the central position of the galaxy (in two
spatial and one frequency dimension), the position angle of the
maximum rotation, the central intensity and scale height of the
emission, the maximum velocity of rotation, and the dispersion along
the line of sight. Besides these eight parameters, the disk models also
need an inclination, and disk models B and C require a scale radius
for the velocity. Finally, the jet model D requires three additional
angles to describe the jet angle and opening angle of the jet, as well
as a terminal radius at which the jet stops accelerating/ionizing the ISM.

From the fits shown in Figure~\ref{fig:models}, it is clear that the
kinematics are inconsistent with a simple rotating disk with a constant velocity (Disk A), which is typically
assumed in order to derive the dynamical masses of marginally resolved
$z\sim6$ quasar host galaxies \citep{wan13,ven16,sha17}.
To account for the extended emission
detected at the systemic velocity, we need a truncated disk model (Disk C) whereby the systemic velocity
goes to zero beyond a certain radius. Such a scenario could arise from
either a gravitationally-dominated disk that has formed in the center
of a dispersion-dominated source, or alternatively, a warp of the
outer disk into the plane of the sky. Both scenarios, however, do not provide adequate fits to the observed \cii\ distribution and kinematics (Figures~\ref{fig:pv} and \ref{fig:models}).

As can be seen in Figure~\ref{fig:pv}, from the position of the black hole, the velocity increases
to a radius of 0.5\,kpc (0\farcs1)
and reaches a peak line-of-sight velocity of 150--200\,km\,s$^{-1}$. Assuming these kinematics can be described by a rotating disk, and the dynamical mass within this radius is
$M_\mathrm{dyn}=2.3\times10^9$\,/\,sin$^2$($i$)\,$M_\odot$, with $i$ the inclination angle of the disk. For comparison,
the mass of the black hole is $1\times10^9$\,$M_\odot$, and the
inferred gas mass within this radius is $1.6\times10^{10}$\,$M_\odot$. For an inclination angle of $i=35^{\circ}$ (see Table~\ref{tab:model}), the dynamical mass and molecular gas mass within the central 1\,kpc are consistent with each other within the large uncertainties. 
Beyond this radius, the line-of-sight velocity is decreasing faster than
Keplerian (shown by the Disk B model in Figure~\ref{fig:pv}), and reaches approximately systemic velocity at the last measured points at 1.5--2.0\,kpc. To summarize, the
kinematics appear to be dispersion-dominated with some overall rotation (i.e.\ net angular momentum) in the central kiloparsec. This implies that most of the gas has not yet settled in a disk.

\section{Origin of the \cii\ cavities}
\label{sec:cavities}

Instead of a kinematic origin, the presence of the two-kiloparsec-scale
\cii\ cavities could be the result of energy injection into
the ISM. Similar shell-like structures are seen in the
neutral ISM of local galaxies \citep[e.g.,][]{wal99}, and are
often interpreted as the results of supernova or AGN feedback. We can calculate
the energy needed to created such structures seen here using \citep{che74} 

\begin{equation}
    E = 5.3\times10^{43}\,n_0^{1.12}\,(d/2)^{3.12}\, v_\mathrm{exp}^{1.4}\,\mathrm{ergs}
\end{equation}

\noindent
with $n_0$ being the density in cm$^{-3}$, $d$ being the diameter of the cavity in parsec, and $v_\mathrm{exp}$ the expansion velocity in \kms. The cavities have a diameter of approximately $\sim$0.5\,kpc
(see Figure~\ref{fig:channel}), and the luminosity-weighted, average density of the ISM in the
quasar host has been estimated to be 10$^5$\,cm$^{-3}$ \citep{ven17b} based on a
multiline analysis in this source. Under the assumption that the
S-like structure is due to energy input in the plane of the galaxy,
the expansion velocity can be approximated by the amplitude of the
derivation from systemic velocity \citep[e.g.,][]{wal99}. We estimate an amplitude of $\sim$80\,\kms\ at the center of the cavities (Figure~\ref{fig:pv}). Taken together, we
derive a required energy of $\sim$$3\times10^{59}$\,erg, or the equivalent of the
mechanical energy output of 300 million typical type II SNe \citep{woo86}. This large number of
supernovae would have to be centrally clustered in both cavities or in between the cavities. If
we assume that 1 supernova explodes for every 200 solar masses of new stars \citep[e.g.,][]{die06}, we
expect 7 supernovae to form each year for the observed star formation
rate of 1500\,\msunyr\ \citep{ven16} . The above total number of supernovae would then require a constant star
formation rate over $\sim$40 million years in a very restricted volume. However, from the dust continuum map (Figure~\ref{fig:moments}) it is clear that the high star-formation rate is distributed over a large, 5\,kpc$\times$5\,kpc area, requiring an even longer period of high star-formation activity in or between the cavities to inject enough energy into the ISM to create the cavities. On the other hand,
the required energy can be easily produced by the central accreting
supermassive black hole which has a bolometric luminosity of $L_\mathrm{bol}=7.5\times10^{46}$\,erg\,s$^{-1}$ \citep{maz17b}, or $2.4\times10^{54}$\,erg\,yr$^{-1}$. Assuming that 5\% of the accreted energy is deposited as thermal energy in the ISM \citep[e.g.,][]{dim05}, the required energy to create the cavities would be generated in only 2.5 million years, which is much shorter than the Salpeter time (or e-folding time) of the black hole of 45 million years \citep[e.g.,][]{der14,maz17b}. If the black hole was indeed the cause of the cavities, the implied energy injection would be toward both sides, consistent with a simple
jet-driven picture. This concept that holes in the gas distribution can be created by AGN jets is supported by numerical simulations \citep[e.g.,][]{gai12}. However, modeling of such a jet requires again a
very unique distribution of the gas (Section~\ref{sec:modelresults}). Regardless, the process needed to create these cavities does not suppress the star formation. This has also been seen in luminous quasars at $z\sim2$ \citep[e.g.,][]{har12} and predicted by some simulations \citep[e.g.,][]{gai12}.

\section{Companion Galaxies}
\label{sec:companions}

\begin{deluxetable*}{lccccc}
\tablecaption{Measured and Derived Properties of the Quasar and the Companion Galaxies. \label{tab:companions}}
\tablewidth{0pt}
\tablehead{\colhead{Source:} & \colhead{Quasar} & \colhead{C1} & \colhead{C2} & 
\colhead{C3} & \colhead{LAE}}
\startdata
R.A. & 03$^\mathrm{h}$05$^\mathrm{m}$16$^\mathrm{s}\!$.919 & 
03$^\mathrm{h}$05$^\mathrm{m}$16$^\mathrm{s}\!$.945 & 
03$^\mathrm{h}$05$^\mathrm{m}$16$^\mathrm{s}\!$.868 & 
03$^\mathrm{h}$05$^\mathrm{m}$16$^\mathrm{s}\!$.381 & 
03$^\mathrm{h}$05$^\mathrm{m}$16$^\mathrm{s}\!$.803 \\
Decl. & --31$^\circ$50$^\prime$55\farcs901 &
--31$^\circ$50$^\prime$55\farcs728 &
--31$^\circ$50$^\prime$55\farcs220 &
--31$^\circ$50$^\prime$54\farcs980 &
--31$^\circ$50$^\prime$57\farcs300 \\
Redshift\tablenotemark{a} & 6.61391$\pm$0.00015 & 
6.6231$\pm$0.0003 &
6.6104$\pm$0.0004 & 
6.6066$\pm$0.0006 &
6.629$\pm$0.001 \\
Distance (kpc) & ... &
2.0 & 
5.1 &
37.4 &
11.0 \\
$\Delta$$v_\mathrm{los}$\tablenotemark{b} (\kms) & ... & 
+361 & 
--137 & 
--289 & 
+595 \\
$F_\mathrm{[CII]}$\tablenotemark{a} (Jy\,\kms) & 5.43$\pm$0.33 &
0.43$\pm$0.05 &
0.31$\pm$0.08 & 
1.13$\pm$0.17 &
$<$0.23 \\
FWHM$_\mathrm{[CII]}$ (\kms) & 225$\pm$15 &
180$\pm$25 & 
135$\pm$40 &
330$\pm$55 & 
-- \\ 
$S_{\mathrm{cont,158\,{\mu}m}}$\tablenotemark{c}  (mJy) & 5.34$\pm$0.19 &
...\tablenotemark{d} & 
$<$0.23 & 
0.58$\pm$0.12 & 
$<$0.33 \\
$L_\mathrm{[CII]}$ (\lsun) & $(5.9\pm0.4)\times10^9$ &
$(4.7\pm0.5)\times10^8$ &
$(3.4\pm0.8)\times10^8$ &
$(1.2\pm0.2)\times10^9$ &
$<2.5\times10^8$ \\
$L_\mathrm{FIR}$ (\lsun) & $(1.60\pm0.06)\times10^{13}$ & 
... &
$<$$5.2\times10^{11}$ &
$(1.3\pm0.3)\times10^{12}$ &
$<$$7.7\times10^{11}$  \\
SFR$_\mathrm{[CII]}$\tablenotemark{e} (\msunyr) & 1016$\pm$73 & 
51$\pm$9 &
35$\pm$10 &
159$\pm$28 & 
$<$25   \\ 
$M_{\mathrm{H}_2}$\tablenotemark{f} (\msun) & $(1.8\pm0.1)\times10^{11}$ &
$(1.5\pm0.2)\times10^{10}$ & 
$(1.1\pm0.2)\times10^{10}$ &
$(3.7\pm0.6)\times10^{10}$ &
$<$$7.8\times10^9$ \\
\enddata
{\bf Notes.} All the quoted errors are $1\sigma$ and the upper limits are $3\sigma$.
\tablenotetext{a}{The redshift and \cii\ line flux ($F_\mathrm{[CII]}$) are measured using a Gaussian fit to the \cii\ line for the quasar and the companions C1, C2, and C3 (as shown in Figure~\ref{fig:companions}). The redshift of the Ly$\alpha$ emitter (LAE) is taken from \citet{far17}. }
\tablenotetext{b}{The line-of-sight velocity $v_\mathrm{los}$ is computed using $v_\mathrm{los} = (z_\mathrm{companion}-z_\mathrm{quasar})/(1+z_\mathrm{quasar})\times c$.}
\tablenotetext{c}{ The continuum flux density at a rest-frame wavelength of 158\,$\mu$m ($S_\mathrm{cont,158\,{\mu}m}$) is measured by averaging the line-free channels in the spectrum around the \cii\ line.}
\tablenotetext{d}{The continuum flux density of C1 could not be accurately determined due to contamination by the quasar host galaxy.}
\tablenotetext{e}{The star-formation rate SFR$_\mathrm{[CII]}$ is derived using SFR$_\mathrm{[CII]} = 3\times10^{-9} L_\mathrm{[CII]}^{1.18}$ \citep{del14}.}
\tablenotetext{f}{Molecular gas mass, derived from the \cii\ luminosity using a \lcii-to-H$_2$ conversion factor of $\alpha_\mathrm{[CII]} = 31$\,\msun/\lsun\ \citep{zan18}.}
\end{deluxetable*}

The environment of J0305--3150 has been the subject of various studies. A study searching for Ly$\alpha$ emission from galaxies at the redshift of the quasar revealed
the presence of a Ly$\alpha$ halo around the quasar, and a faint
Ly$\alpha$ emitting (LAE) companion at a distance of
12.5\,kpc \citep{far17}. In addition, a larger field-of-view narrowband
search for Ly$\alpha$ emitters around the quasar indicated that the
quasar is not located in a dense environment on megaparsec
scales \citep{ota18}. Our previous, shallow ALMA data of the field contained two faint (rest frame 158\,$\mu$m flux densities of $S_{\mathrm{cont,158\,{\mu}m}} \ll 1$\,mJy) continuum sources \citep{ven16}. 

In addition to high-resolution imaging of the quasar host galaxy, the
improved sensitivity of the new ALMA data allowed us to search for additional sources in the field. Previously, bright companion sources have been found in the vicinity of some high-redshift quasars \citep[e.g.,][]{dec17,tra17}.
Our new data reveal the presence of three \cii\ emitters at the same redshift as the quasar
(Figure~\ref{fig:companions}; physical parameters in Table~\ref{tab:companions}). One companion, C1, could already be seen in Figure~\ref{fig:channel}. The brightest companion, C3, was already identified as a continuum source in the field in \citet{ven16}. 

The FIR luminosity of companion C3 is comparable to that of the \cii\ companions near several $z\sim6$ quasars presented in \citet{dec17}. However, in contrast to the companions found by \citet{dec17}, all companions identified near J0305--3150 have FIR luminosities that are a factor $\lesssim$10 smaller than that of the quasar host. Also, the estimated molecular gas masses of the companions are at least a factor of $\sim$5 smaller than that of the quasar host. Despite the small masses, the interaction of companions with the quasar host could provide an alternative explanation of the gas morphology and kinematics in the quasar host galaxy. Furthermore, it is a possibility that such interactions triggered the high star-formation rate in the quasar host and the AGN activity. 

\section{Summary}
\label{sec:summary}

The high spatial resolution ($\sim$400\,pc) imaging of a quasar host galaxy at $z=6.6$ shows that its formation is a complex and chaotic process. We find that the ISM in the quasar host has not yet settled in a simple disk. While there are signs that the AGN is affecting the ISM, this feedback is not suppressing the formation of stars in the quasar host. In fact, the star-formation rate inferred by the FIR luminosity of $\sim$1500\,\msunyr\ \citep{ven16} is high among galaxies and quasar hosts at similar redshifts \citep[e.g.,][]{car13,ven18}. The observed high star-formation rate and rapid black hole growth could be triggered by interactions with the newly detected nearby companion galaxies. The observations presented here show the unique role that high-angular resolution observations with ALMA can play in studies of the ISM in some of the most distant massive galaxies.

\vspace{-0.25cm}
\acknowledgments

B.P.V., M.N., and F.W.\ acknowledge funding through the ERC grant ``Cosmic Gas." 
This paper makes use of the following ALMA data:
ADS/JAO.ALMA\#2017.1.01532.S. ALMA is a partnership of ESO
(representing its member states), NSF (USA) and NINS (Japan), together
with NRC (Canada), and NSC and ASIAA (Taiwan), in cooperation with the
Republic of Chile. The Joint ALMA Observatory is operated by ESO,
AUI/NRAO, and NAOJ.

\facilities{ALMA}

\end{document}